\documentclass[preprint,12pt]{elsarticle}

\usepackage{amssymb}

\usepackage{amsmath}
\usepackage{subfigure}
\usepackage{gensymb}
\usepackage{upgreek}
\usepackage{bm}
\usepackage{subfigure}
\usepackage{booktabs}

\usepackage{url}

\usepackage{breakurl}

\begin{document}

\begin{frontmatter}



\title{Efficient representation of head-related transfer functions in continuous space-frequency domains}


\author[1]{Adam Szwajcowski}
\ead{szwajcowski@agh.edu.pl}

\address[1]{AGH University of Science and Technology, Department of Robotics and Mechatronics}

\begin{abstract}
Utilizing spherical harmonic (SH) domain has been established as the default method of obtaining continuity over space in head-related transfer functions (HRTFs). This paper concerns different variants of extending this solution by replacing SHs with four-dimensional (4D) continuous functional models in which frequency is imagined as another physical dimension. Recently developed hyperspherical harmonic (HSH) representation is compared with models defined in spherindrical coordinate system by merging SHs with one-dimensional basis functions. The efficiency of both approaches is evaluated based on the reproduction errors for individual HRTFs from HUTUBS database, including detailed analysis of its dependency on chosen orders of approximation in frequency and space. Employing continuous functional models defined in 4D coordinate systems allows HRTF magnitude spectra to be expressed as a small set of coefficients which can be decoded back into values at any direction and frequency. The best performance was noted for HSHs and SHs merged with reverse Fourier-Bessel series, with the former featuring better compression abilities, achieving slightly higher accuracy for low number of coefficients. The presented models can serve multiple purposes, such as interpolation, compression or parametrization for machine learning applications, and can be applied not only to HRTFs but also to other types of directivity functions, e.g. sound source directivity.

\end{abstract}



\begin{keyword}
HRTF \sep spherical harmonics \sep hyperspherical harmonics \sep continuous functional model



\end{keyword}

\end{frontmatter}


\section{Introduction}

With the constant development of virtual and augmented reality, the need for binaural rendering is increasing year by year. A key component to immersive sound reproduction are head-related transfer functions (HRTFs), which describe natural filters of one's head, pinna and torso depending on the direction of incoming sound. The spectral cue provided by HRTF magnitudes is particularly important to sound localization on sagittal planes, where binaural cues such as interaural level difference (ILD) and interaural time difference (ITD) are invariant. Since they depend on fine details of one's morphology, they are different for every individual; application of individually measured HRTFs has been proven to significantly improve sound localization in numerous studies (e.g. \citep{Wenzel1993,Katz2012,Parseihian2012,Stitt2019}).

HRTFs are continuous functions of direction, distance and frequency. However, due to technological limits, they can only be measured at discrete points in space and time. Usually, they are stored as sets of head-related impulse responses (HRIRs) corresponding to certain directions, from which HRTFs can be then obtained by means of Fourier transform.

HRTFs (or HRIRs) are relatively large and complex datasets, usually comprising of hundreds of thousands of samples. This precision is excessive, since humans have been proven to be insensitive to fine spectral details (e.g. \citep{Kulkarni1998,Breebaart2001,Xie2010}). For this reason, numerous models have been developed to simplify HRTFs without crippling their perceptual relevance. While some of them focused purely on data compression (e.g. \citep{Kistler1992,Wang2009,Shekarchi2013,Arevalo2020}), others aimed to also reclaim the underlying continuity lost by the measurement constraints. Within the latter, three groups can be distinguished.

The first group concerned approximating the spectra. In particular, in early years, different variants of modeling HRTFs as filters of infinite impulse response were investigated by many researchers\footnote{Finite-impulse-response filters were also studied (e.g. \citep{Kulkarni1995,Wu1997}), but they do not provide continuity over frequency.} \citep{Blommer1997,Huopaniemi1999,Kulkarni2004}. Following that trend, Ramos and Cobos proposed an approach based on a set of parametric filters, commonly used in audio equalization \citep{Ramos2013}. Recently, Szwajcowski evaluated the possibility of modeling HRTF spectra by means of common basis functions \citep{SzwajcowskiICSV}.

The second group includes models which provide continuity in space. One of the first attempts was made by Chen et al. in 1995, who approximated HRTFs with thin-plate splines \citep{Chen1995}. In 1998, Evans et al. presented the possibility of expressing the space dependency of HRTFs by means of spherical harmonics (SHs) \citep{Evans1998}. Although other space-continuous HRTF models have also been proposed since then (e.g. utilizing Slepian functions or spherical wavelets \citep{Bates2015,Hu2019}), the SH representation is currently still the only standardized way of storing HRTFs other than discrete data \citep{SOFA2020}. However, there is still ongoing research comparing different approaches to the SH approximation, evaluating various preprocessing techniques or using mixed-order SHs \citep{Brinkmann2018,Li2021,Engel2022}.

Finally, a few multidimensional representations have also been developed, covering both spectral and spatial dependencies. In 2008, Abhayapala described a theoretical model based on SHs for direction and spherical Bessel functions for frequency \citep{Abhayapala2008}. In 2010, W. Zhang et al. extended this model utilizing spherical Hankel functions for distance\footnote{The spherical Hankel functions were first used for describing the distance-dependent HRTFs by Duraiswami et al. in 2004 \citep{Duraiswami2004}.} and evaluated it on exemplary datasets \citep{Zhang2010}. In 2015, M. Zhang et al. proposed a similar representation but with spherical Bessel functions replaced by complex exponentials \citep{Zhang2015}. In both studies, coefficients were determined by least-squares fitting followed by numerical integration, effectively handling the computations separately for each dimension. Recently, Szwajcowski presented a model based on hyperspherical harmonics (HSHs), in which multidimensional basis functions were fit by means of least-squares within a single matrix equation \citep{SzwajcowskiHSH}. While this approach requires much more computational power, it is also more accurate and reflects coupling between space and frequency inherent for underlying directivity functions. Despite lack of physical motivation to express HRTFs as hyperspherical functions, the model featured relatively low reproduction errors, even outperforming discrete compression methods.

The introduction of HSHs presented a whole new method of developing HRTF representations, but it also raised a couple of questions regarding that approach:

\begin{enumerate}
	\item{Is the HSH representation the most efficient one of this kind? How do other 4D models compare?}
	\item{What is the impact of changing resolution in space and frequency on the reproduction accuracy of the analyzed models?}
	\item{What is the optimal range of values of the parameters controlling the resolution in space and frequency?}
\end{enumerate}
	
\noindent This paper aims to find answers to all these questions and provide a robust comparison of such models. They hold several notable advantages over the SH representation. One of them is continuity over frequency, which allows to extract HRTF magnitude not only at any given direction but also at any given frequency without resampling or other kinds of interpolation, thus being computationally attractive. Furthermore, they allow for an easy control between accuracy and amount of data by adjusting the approximation parameters; acknowledging psychoacoustical aspects in the process of deriving the functional model can lead to significant reduction of data size by ignoring high-order basis functions responsible for imperceptible spectral details. Finally, 4D representations can be useful in machine learning applications, as they capture complete information on spectral and spatial variability within a single set of coefficients.

Section \ref{s:theory} lays down necessary theoretical background on 4D coordinate systems and basis functions defined within them. Section \ref{s:methods} describes the implementation and evaluation of the HRTF representations employed within this research. Section \ref{s:comparison} presents results of basic comparison between the chosen models. Section \ref{s:optimal} focuses on the relation between accuracy and number of coefficients. The results are discussed in section \ref{s:discussion}, while section \ref{s:conclusions} concludes the work.

\section{Theoretical background}
\label{s:theory}

\subsection{4D coordinate systems}
\label{s:coordinates}

This research concerns models relying on SHs for the spatial dependency and on another dimension for the spectral dependency. Thus, 4D coordinate systems need to be constructed by extending the spherical coordinate system. Depending on the nature of the extra dimension, the result is either spherindrical coordinate system (SCS) or hyperspherical coordinate system (HCS)\footnote{HCS can be defined for any number of dimensions above 3, but within this work this acronym refers only to the 4D HCS.} for linear and angular dimension, respectively. Assuming spherical coordinate system defined as ($r$,$\varphi$,$\theta$), where $r$ is distance, $\varphi \in [0,2\pi)$ is azimuth and $\theta \in [0,\pi]$ is inclination, SCS can be simply defined as ($r$,$\varphi$,$\theta$,$w$), with $w$ being an extra linear dimension (later replaced by $f$, since within this work that extra dimension is reserved for frequency). On the other hand, HCS includes an additional angle $\psi \in [0,\pi]$, which can be utilized for expressing frequency. The relation between HCS and 4D Cartesian system ($x$,$y$,$z$,$w$) is following:

\begin{equation}
\begin{aligned}
	&x = \rho \sin{\psi} \sin{\theta} \sin{\varphi} \, ,\\
	&y = \rho \sin{\psi} \sin{\theta} \cos{\varphi} \, ,\\
	&z = \rho \sin{\psi} \cos{\theta} \, ,\\
	&w = \rho \cos{\psi} \, ,
\label{e:HCS}
\end{aligned}
\end{equation}

\noindent where $\rho$ is hyperspherical radius satisfying the equation $\rho^{2} = r^{2} + w^{2}$.

This study concerns only far-field HRTFs, for which the radial dependence can be dropped. The basis functions described in the following subsections thus depend only on $\varphi$ for azimuth, $\theta$ for inclination and either $w$ or $\psi$ for frequency. Since the models are defined in 4D coordinate systems, they are still referred to as 4D, even though they depend on only three variables.

\subsection{Hyperspherical harmonics}
\label{s:HSH}

HSHs are functions defined on the hypersphere (3-sphere). Real HSHs are given by the following equation \citep{Domokos1967}:

\begin{equation}
	Z_{nl}^{m}(\varphi,\theta,\psi) \equiv 2^{l+\frac{1}{2}} (l+1)! \sqrt{\frac{2(n+1)(n-l+1)!}{\pi (n+l+2)!}} \sin^l{\psi} \; C_{n-l}^{l+1}(\cos{\psi}) \; Y_{l}^{m}(\varphi,\theta) \, ,
\label{e:HSH}
\end{equation}

\noindent where $C_{\nu}^{\alpha}(x)$ are Gegenbauer polynomials (also known as ultraspherical polynomials):

\begin{equation}
\begin{aligned}
	C_{0}^{\alpha}(x) = &1 \\
	C_{1}^{\alpha}(x) = &2 \alpha x \\
	C_{\nu}^{\alpha}(x) = &\frac{1}{\nu} \big{(}2x(\nu+\alpha-1)C_{\nu-1}^{\alpha}(x) - (\nu+2\alpha-2)C_{\nu-2}^{\alpha}(x) \big{)} \,  ,
\label{e:gegenbauer}
\end{aligned}
\end{equation}

\noindent and $Y_{l}^{m}(\varphi,\theta)$ are SHs:

\begin{equation}
Y_{l}^{m}(\phi,\theta) \equiv \sqrt{\frac{2l+1}{4 \pi} \frac{(l-|m|)!}{(l+|m|)!}} \times
\begin{cases}
\sqrt{2} P_{l}^{m}(\cos{\theta}) \cos{\left( m\varphi \right)} \, , & \text{if}\ m > 0 \\
P_{l}^{m}(\cos{\theta}) \, , & \text{if}\ m = 0 \\
\sqrt{2} P_{l}^{|m|}(\cos{\theta}) \sin{\left( |m|\varphi \right)} \, , & \text{if}\ m < 0 \\
\end{cases} \, ,
\label{e:SH}
\end{equation}

\noindent where $P_{l}^{m}$ are associated Legendre functions of degree $l$ and order $m$. HSHs are indexed by three integer parameters $n$, $l$ and $m$, which need to satisfy the following inequalities:

\begin{equation}
\begin{aligned}
	&n \geq 0 \\
	0 \leq \, &l \leq n \\
	-l \leq \, &m \leq l \, .
\label{e:nlm}
\end{aligned}
\end{equation}

In practical applications, the number of HSHs has to be limited. The simplest way to do that is to only define $n_{\text{max}}$ (maximum value of $n$) since $l_{\text{max}}$ and $m_{\text{max}}$ are then also indirectly limited. However, one can also impose independent limits for each parameter. The set of inequalities \eqref{e:nlm} then takes the form:

\begin{equation}
\begin{aligned}
	0 \leq &n \leq n_{\text{max}} \\
	0 \leq \, &l \leq \min(n,l_{\text{max}}) \\
	- \min (m_{\text{max}}, l) \leq \, &m \leq \min (m_{\text{max}}, l) \, .
\label{e:nlmlim}
\end{aligned}
\end{equation}

\noindent Such decoupling of limits is particularly useful when utilizing HSHs for HRTF representation, as it allows for independent control of resolution in space and in frequency; $l_{\text{max}}$ and $m_{\text{max}}$ are responsible for resolution over azimuth and elevation, respectively, while $n_{\text{max}}$ controls resolution over $\psi$, i.e. over frequency. While it makes sense to impose $l_{\text{max}}$ = $m_{\text{max}}$, it is recommended for both these limits to be lower than $n_{\text{max}}$ (see section \ref{s:orders} for explanation). 

In order to apply HSHs to directivity functions, frequency must be mapped to the $\psi$ angle. To leverage HCS properties, the mapping is done only on half the hypersphere, i.e. for $\psi \in [0,\pi/2]$. This way, only the 0 frequency (constant component) lies on the hyperpole\footnote{Hyperpoles are points for which all $\varphi$ and $\theta$ converge to the same direction. They occur at $\psi = 0$ and $\psi = \pi$.}, which is desired, as geometrical properties of HCS near hyperpoles reflect the omnidirectionality of directivity functions at low frequencies. Maximum available frequency of an HRTF set is half the sampling frequency ($f_{s}$), which is mapped to the hyperequator ($\psi = \pi/2$). Assuming linear mapping, the relation between $\psi$ and frequency can be portrayed by the formula:

\begin{equation}
\psi(f) = \frac{\pi f}{f_{s}} \, .
\label{e:mapping}
\end{equation}

\noindent As this mapping is performed on only half of the available range of $\psi$, it requires twice as much resolution; however, it also allows all HSHs non-symmetric about the hyperequator to be ignored, which comprise almost half of the basis. This procedure effectively removes the convergence at high frequencies, while increasing required number of functions only ever so slightly. For more detailed description of HSHs and their application to express HRTFs, Reader is advised to read the original paper \citep{SzwajcowskiHSH}.

\subsection{Functions in spherindrical coordinate system}
\label{s:SHF}

While HSHs are an obvious choice for basis functions in HCS, SCS and corresponding bases are much less common in the literature. However, they can be easily created by merging SHs with one-dimensional (1D) basis functions, following previous research regarding multidimensional continuous functional HRTF models \citep{Zhang2010,Zhang2015}. Depending on the study and adopted evaluation methods, different 1D bases were proposed, including various variants of Fourier-Bessel series and complex exponentials \citep{Zhang2010,Zhang2015,Zhang2009,Zhang2009a}. All these works concerned high-resolution complex HRTFs; when investigating only magnitude approximations, both Fourier series and Fourier-Bessel series performed on par for 16+ basis functions, while providing lower errors than polynomials \citep{SzwajcowskiICSV}. Since this work adopts a similar modeling approach, representations incorporating both Fourier and Fourier-Bessel series are investigated and referred to as SH-Fourier-series (SHFS) and SH-Fourier-Bessel-series (SHFBS), respectively.

SHFS is obtained by multiplication of SHs and a variation of Fourier series based on cosines, as used by Kulkarni and Colburn for spectral approximation of HRTFs \citep{Kulkarni1998}. SHFS is given by the formula:

\begin{equation}
\label{e:SHF}
F_{nl}^{m}(\varphi,\theta,f) \equiv Y_{l}^{m}(\varphi,\theta) \, \cos{\left(\pi n\frac{f}{f_{s}}\right)} \, ,
\end{equation}

\noindent where $n$ is order of Fourier series ($n = 0, 1, 2, ...$).

SHFBS has a very similar definition, except Fourier series are replaced by Fourier-Bessel series of order 0 ($J_{0}$). The basis is supplemented by constant component for better comparison with SHFS and HSHs, both of which are definable for $n$ = 0. SHFBS can be thus given as:

\begin{equation}
\label{e:FBS}
B_{nl}^{m}(\varphi,\theta,f) \equiv Y_{l}^{m}(\varphi,\theta) \times
\begin{cases}
1 \, , \, &n = 0 \\
J_{0} \left(\mu_{n} \frac{2 f}{f_{s}} \right) \, , \, &n > 0 \\
\end{cases}
\, ,
\end{equation}

\noindent where $\mu_{n}$ is $n$th root of $J_{0}$.

Bessel functions consist of ripples with decreasing magnitude. When applying to HRTFs, this means lower variability for higher frequencies, while the opposite is desirable. Thus, reversed Fourier-Bessel series is also introduced by mirroring normal Fourier-Bessel functions. SH-reversed-Fourier-Bessel-series (SHRFBS) representation is defined as: 

\begin{equation}
\label{e:RFBS}
\hat{B}_{nl}^{m}(\varphi,\theta,f) \equiv Y_{l}^{m}(\varphi,\theta) \times
\begin{cases}
1 \, , \, &n = 0 \\
J_{0} \left(\mu_{n} \left( 1 - \frac{2 f}{f_{s}} \right) \right) \, , \, &n > 0 \\
\end{cases}
\, .
\end{equation}

Alternatively, other functions or other variants of the ones defined above could be applied, e.g. employing Fourier series of non-integer orders or utilizing different orders of Bessel functions. Since the possibilities are endless, for the sake of simplicity, this study focuses only on the three SCS bases described above.

\section{Methods}
\label{s:methods}

\subsection{Discrete HRIR data to 4D basis function domain}
\label{s:HRIR}

HRTFs are usually stored in time domain, i.e. in the form of HRIRs. Even though time is more intuitive than frequency to be modeled as the fourth dimension, time-based approach to HRTF modeling has been proven to be less accurate than the frequency-based one \citep{Evans1998,Hartung1999}. Furthermore, HRIRs include information on phase, which is widely acknowledged to be irrelevant to localization as long as ITD is preserved \citep{Kistler1992,Kulkarni1999,Romigh2015}, although conflicting results have been presented regarding whether or not the phase linearization is detectable, particularly at low frequencies \citep{Kulkarni2004,Rasumow2014,Andreopoulou2022}. This work focuses only on the magnitude part of HRIR spectra, assuming that the phase can be either linearized basing on ITD or modeled independently.

The goal of the models described within this paper is to accurately express magnitude of HRTFs within a relatively small set of coefficients $\alpha_{nl}^{m}$, from which data can be then decoded for any requested direction and frequency by means of a weighted sum:

\begin{equation}
\begin{aligned}
\hat{H}(\varphi,\theta,f) =  \sum_{n=0}^{n_{\text{max}}} \quad \sum_{l=0}^{\min{(n,l_{\text{max}})}} \sum_{m=-\min{(l,m_{\text{max}})}}^{\min{(l,m_{\text{max}})}} \alpha_{nl}^{m} X_{nl}^{m}(\varphi,\theta,\gamma) \, ,
\end{aligned}
\label{e:sum}
\end{equation}

\noindent where $X_{nl}^{m}$ are 4D basis function ($Z_{nl}^{m}$, $F_{nl}^{m}$, $B_{nl}^{m}$, $\hat{B}_{nl}^{m}$ or any other of similar definition) and $\gamma$ symbolizes the extra dimension (either $\psi(f)$ for HCS or $f$ for SCS). In the past, multidimensional HRTF models employed a hybrid method to determining $\alpha_{nl}^{m}$; first, SH coefficients were computed by means of least-squares fitting for each frequency bin separately, and then direct integration over the SH coefficients was applied to include the frequency dependence \citep{Zhang2010,Zhang2015}. While direct integration and least squares converge to the same solution, the latter yields lower approximation errors when the basis is truncated. This study follows the approach from the recent work concerning the HSH representation, in which the coefficients are determined by a single, large least-squares fitting \citep{SzwajcowskiHSH}. This fitting includes all the spectro-spatial variability of a given HRTF set and can be described by the following matrix equation:

\begin{equation}
\label{e:matrixLS}
\begin{bmatrix}
X_{00}^{0}(\Omega_{1}) & \dots & X_{n_{\text{max}}l_{max}}^{m_{\text{max}}}(\Omega_{1}) \\
\vdots & \ddots & \vdots \\
X_{00}^{0}(\Omega_{K}) & \dots & X_{n_{\text{max}}l_{\text{max}}}^{m_{\text{max}}}(\Omega_{K}) \\
\end{bmatrix}
\begin{bmatrix}
\alpha_{00}^{0} \\ \vdots \\ \alpha_{n_{\text{max}}l_{\text{max}}}^{m_{\text{max}}}
\end{bmatrix} \\
 =
\begin{bmatrix}
H(\Omega_{1}) \\ \vdots \\ H(\Omega_{K})
\end{bmatrix} \, ,
\end{equation}
\vspace{5mm}

\noindent where $H$ denotes the original HRTF values and $\Omega_{k}$ is frequency-direciton triplet of $k$th data point:

\begin{equation}
\label{e:Omega}
\Omega_{k} \equiv (\varphi_{k}, \theta_{k}, \gamma_{k}) \, .
\end{equation}

Furthermore, weights are included in the least-squares computation to minimize the impact of data lying in perceptually irrelevant regions. Equation \eqref{e:matrixLS} then takes the form of:

\begin{equation}
\label{e:weightedLS}
(\mathbf{X^{\mathrm{T}} W X) \bm{\upalpha} = X^{\mathrm{T}} W H} \, ,
\end{equation}

\noindent where $\mathbf{X}$, $\bm{\upalpha}$, and $\mathbf{H}$ denote the respective matrices from Eq. \eqref{e:matrixLS}, $^{\mathrm{T}}$ denotes matrix transposition and $\mathbf{W}$ is a diagonal matrix with weights for consecutive frequency-direction triplets:

\begin{equation}
\label{e:weightMatrix}
\mathbf{W} =
\begin{bmatrix}
w(\Omega_{1}) & 0 & \dots & 0 \\
0 & w(\Omega_{2}) & & \vdots \\
\vdots & & \ddots & 0 \\
0 & \dots & 0 & w(\Omega_{K}) \\
\end{bmatrix} \, ,
\end{equation}
\vspace{5mm}

\noindent where $w$ has lower values outside of human frequency hearing range:

\begin{equation}
\label{e:weight}
w(\Omega) =
\begin{cases}
0 \, , \, &f < f_{l} \\
1 \, , \, &f_{l} < f <  f_{u} \\
\cos{ \left( \frac{2(f-f_{u})}{f_{s} - f_{u}} \pi \right) } \, , \,  &f > f_{u}
\end{cases}
\, ,
\end{equation}

\noindent where $f_{l}$ and $f_{u}$ are lower and upper frequency limits, assumed to be 20 and 20 000 Hz, respectively. By gradually decreasing weights above 20 kHz, one can avoid large empty regions, which could cause overfitting. On the other end of the scale, no such issue exists, as there is usually only one sample below 20 Hz representing the constant component. In some cases, a regularization or preapproximating some datapoints might be needed anyway, since HRTF sets often lack data for low elevations \citep{Zotkin2009,Ahrens2012}. What is more, for equiangular sampling on the sphere, Eq. \eqref {e:weight} should also include area-weighting \citep{Szwajcowski2021}. However, the dataset used in this work features a sampling scheme which requires neither of these techniques (details in section \ref{s:evaluation}).

In the literature, two conflicting methods of processing HRTF magnitude can be found; some researchers propose modeling in the linear scale, which minimizes the differences in energy (e.g. \citep{Engel2022,Zotkin2009,Ben-Hur2019}), while others suggest using the logarithmic scale, arguing that it is closer to how the human auditory system perceives sound (e.g. \citep{Kulkarni2004,Li2021,Romigh2015}). While the latter appears to be more popular in recent research, there is no clear evidence that it is indeed superior. Brinkmann and Weinzierl compared different types of HRTF preprocessing and both linear and logarithmic magnitude achieved similar results according to binaural models \citep{Brinkmann2018}. The former also required lower maximum order of SHs to achieve the energy criterion defined by the authors, which was expected since the energy is defined in the linear scale as well. Recent study by Szwajcowski suggests that HRTF distance metrics defined in the linear scale have stronger correlation with subjective ratings, although the differences are not large \citep{SzwajcowskiICA}. Thus, the linear scale was chosen for this work.

All the computations were performed in \textsc{Matlab} using \textsc{ooDir} Toolbox \citep{ooDir}. The 4D HRTF models can be found in its database to enhance their reproduction and comparison with other HRTF representations.

\subsection{Space and frequency approximation orders}
\label{s:orders}

\begin{figure}
\centering
\includegraphics[scale=0.6]{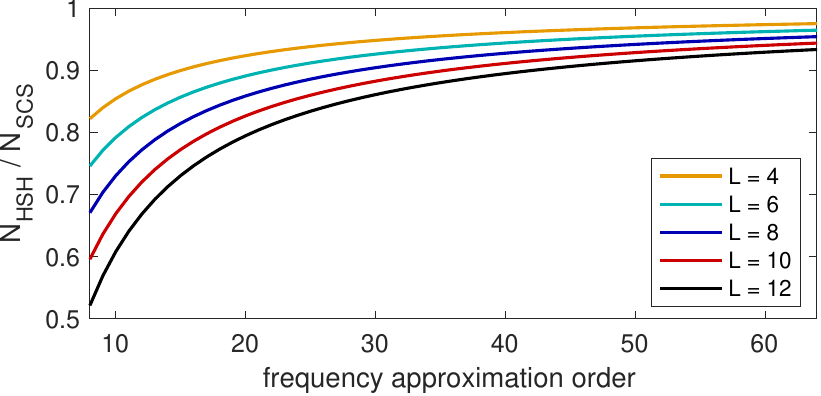}
\caption{Coefficient number ratio for HSHs and functions in SCS for the same max SH and frequency approximation orders}
\label{f:N}
\end{figure}

All investigated functions ($Z_{nl}^{m}$, $F_{nl}^{m}$, $B_{nl}^{m}$ and $\hat{B}_{nl}^{m}$) are described by three parameters: $l$ and $m$ correspond to degree and order of embedded SHs, respectively, while $n$ determines the resolution over the frequency axis. For the sake of convenience, maximum values of $l$ and $m$ are referred to as \textit{max SH order}\footnote{Within this work $l_{\text{max}} = m_{\text{max}}$ is always assumed.} and denoted as $L$. For the frequency part, the term \textit{frequency approximation order} denoted by $\eta$ is introduced, where for SHFS, SHFBS and SHRFBS $\eta = n_{\text{max}}$, while for HSH $\eta = n_{\text{max}}/2$ (see section \ref{s:HSH} for explanation). This way, models in both HCS and SCS have the same resolution in space and frequency for the same values of $L$ and $\eta$; however, the bases do not feature the same number of functions even when their max SH order and frequency approximation order are equal. For the SCS bases, the number of functions is given simply as:

\begin{equation}
\label{e:N_SCS}
N_{\text{SCS}}(L,\eta) = (L+1)^{2}(\eta+1) \, ,
\end{equation}

\noindent while for HSHs it is more complex due to the definition of the basis and the fact that the HSHs which are not symmetric about the hyperequator are ignored. The number of HSHs for a given $L$ and $\eta$ is:

\begin{equation}
\label{e:N_HSH}
N_{\text{HSH}}(L,\eta) =  \sum_{n=0}^{2\eta} \sum_{l=0}^{\min{(n,L)}} \sum_{m=-l}^{l} 1 - (n-l) \, \% \, 2 \, ,
\end{equation}

\noindent where \% denotes modulo operation, i.e. the sum increases only when $n-l$ is even (HSHs for odd $n-l$ do not satisfy the symmetry criterion). For the same approximation orders, $N_{\text{HSH}}$ is smaller than $N_{\text{SCS}}$ and their relation depends on $L$ and $\eta$ (Fig.~\ref{f:N}). Even though, for the same $L$ and $\eta$, functions defined in both HCS and SCS technically have the same maximum number of zero-crossings along respective dimensions, the lower number of functions for HSHs results from the basis being formed in such a way, that for a given $n_{\text{max}}$ the resolution is spread evenly on the hypersphere. When HRTF data is mapped to a half of the hypersphere, the low frequency datapoints lie closer to each other (near hyperpole) than the high frequency ones (near the hyperequator). This way, some of the spatial resolution at lower frequencies is being sacrificed, while maintaining the continuity in frequency.

While approximation in both space and frequency at the same time is novel, some research have been carried out for perceptual impact of approximation in either of these dimensions independently. For frequency, Kulkarni and Colburn showed that HRTF approximation with Fourier series of $\eta = 16$ is on the brink of being consistently recognized, while for frequency approximation order of 32 or higher, subjects performed at chance\footnote{Due to ambiguous description, actual $\eta$ might be lower by 1.} \citep{Kulkarni1998}. Although the experiment design was quite simplistic, Senova et. al obtained consistent results with more sophisticated evaluation methods \citep{Senova2002}. Breebaart et al. investigated parametrizing HRTFs on ERB scale, where for most positions around 20 parameters were needed to achieve perceptual irrelevance of the approximation \citep{Breebaart2010}. On the other hand, Andreopoulou and Katz reported results, in which minimum-phase HRTFs for $\eta = 32$ were clearly discernible from full-phase 256-sample data \citep{Andreopoulou2022}. This discrepancy might be caused by using prescreening, which ensured that the subjects were very competent for the task, with above-average hearing abilities.

As far as the resolution in the SH domain is concerned, different formulae have been proposed for max SH order capable of capturing all the HRTF variability for a given frequency \citep{Zhang2010,Zhang2015}. These formulae suggest orders up to 30 or 46, which are not viable for most of measured HRTF datasets as they are usually relatively sparsely sampled. On the other hand, Romigh et al. showed that some localization cues already occur for $L = 2$, and for $L = 6$ the SH representation (based on logarithmic magnitude and accompanying ITD values) allowed to localize sound sources equally well as for the full HRTF set \citep{Romigh2015}. However, the research did not investigate other perceptual effects such as timbre, and so higher max SH orders might be needed to actually make the rendered sound indiscernible. Li et. al did take these effect into account and in their mixed-order approach a combination of $L = 8$ for low and $L = 18$ for high frequencies proved to be accurate enough to be perceived the same as the signals obtained with the original HRTFs \citep{Li2021}.

Within this work, different orders of approximation in space and frequency will be evaluated. The above-cited research will serve as a basis for estimating the range of analysis in the following sections.

\subsection{Evaluation method}
\label{s:evaluation}

In order to estimate the performance of the 4D models in practice, the evaluation was performed on individual HRTFs from the HUTUBS database \citep{Brinkmann2019}. The HRTFs were measured at 10{\degree} resolution in elevation and resolution in azimuth varying so that the distance between the points was roughly the same for all elevations, yielding 440 measurement points in total, which cover the entire sphere. Even though the database includes data for 96 subjects, previous work showed that performing computations on the entire database is excessive, as variability in approximation accuracy for different basis functions is relatively low \citep{SzwajcowskiICSV}. To reduce the computational cost, the evaluation was performed on only 10 HRTF sets. The HRTFs were chosen randomly from the database and their IDs are: 3, 6, 22, 32, 34, 35, 40, 55, 91 and 92.

Alike least-squares fitting, depending on the study, the accuracy is also evaluated in either linear or in logarithmic scale, or even in both \citep{Hugeng2017}. Consequently, this work employs mean-square-error (MSE), which is defined in the linear scale and which has been shown to be a relatively reliable predictor of perceived differences \citep{SzwajcowskiICA}. MSE is given by the following formula\footnote{Least-squares fitting in the linear scale can result in negative magnitude in some regions; since negative magnitude has no physical interpretation, the value of datapoints lying in such regions is assumed to be 0.}:

\begin{equation}
\text{MSE} = \frac{\sum\limits_{k} |\hat{H}(\Omega_{k}) - H(\Omega_{k})|^{2}}{\sum\limits_{k}|H(\Omega_{k})|^{2}} \times 100\% \, ,
\label{e:MSE}
\end{equation}

\noindent where the range of summing depends on the type of analysis; for analysis in frequency, the averaging is performed over the space, for analysis in space, the averaging is performed over the frequency, and, for a single-value error, the averaging is performed over all the samples. Datapoints whose center frequencies lie outside of the human hearing range (assumed 20 - 20 000~Hz) are excluded from the calculation. In each case, MSE is determined separately for HRTFs for both ears of a given subject and then averaged. Thus, averaging over 10 subjects is in fact averaging over 20 different HRTF sets (two HRTF sets per subject).

\section{Model comparison}
\label{s:comparison}

In order to determine which of the bases is the best fit for representing HRTFs, the accuracy was compared for several exemplary configurations of $L$ and $\eta$. To maintain the possibility of reference to previously cited perceptual-oriented research, $\eta$ was set to 16, 32 or 64. These values were paired with $L \in \{4,6,9\}$; such values of $L$ result in 25, 49 and 100 SHs, respectively, following the pattern of roughly doubling the number of functions with consecutive orders, while also staying in range of typical SH-related research (e.g. \citep{Romigh2015}). The combinations of certain approximation orders are denoted  as $L$:$\eta$, e.g. 4:16 means a configuration with $L=4$ and $\eta=16$.

Due to the difference between number of functions for HSHs and SCS bases, the comparison includes HSHs with increased $\eta$ to match the number of functions with the other analyzed bases. These matched HSH configurations are denoted as HSH$_{\text{m}}$. Table \ref{t:orders} lists all configurations included in the model comparison.

\begin{table}[h]
\caption{List of configurations of $l$, $\eta$ and $\eta_{m}$ (true $\eta$ for HSH$_{m}$) and corresponding numbers of functions. Values in parentheses indicate relative difference between $N_{\text{HSH}}$ or $N_{\text{HSH$_{m}$}}$ and $N_{\text{SCS}}$, respectively. Please note that $\eta_{m}$ can have half values, since for HSHs $\eta = n_{\text{max}}/2$.}
\vspace{2mm}
\centering
\begin{tabular}{@{}cccccc@{}}
\toprule
$L$ & $\eta$ & $\eta_{m}$ & $N_{\text{SCS}}$ & $N_{\text{HSH}}$ & $N_{\text{HSH$_{m}$}}$ \\ \midrule
4   & 16     & 17.5       & 425               & 385 ($-9.4\%$)                                              & 420 ($-1.2\%$)                                                   \\
4   & 32     & 33.5       & 825               & 785 ($-4.8\%$)                                              & 820 ($-0.6\%$)                                                   \\
4   & 64     & 65.5       & 1625             & 1585 ($-2.5\%$)                                            & 1620 ($-0.3\%$)                                                 \\
6   & 16     & 18.5       & 833               & 721 ($-12.6\%$)                                            & 840 ($+0.8\%$)                                                  \\
6   & 32     & 34.5       & 1617             & 1505 ($-6.9 \%$)                                           & 1624 ($+0.4\%$)                                                \\
6   & 64     & 66.5       & 3185             & 3073 ($-3.5\%$)                                            & 3192 ($+0.2\%$)                                                \\
9   & 16     & 19.5       & 1700             & 1365 ($-19.6\%$)                                          & 1720 ($+1.2\%$)                                                \\
9   & 32     & 35.5       & 3300             & 2965 ($-10.2\%$)                                          & 3320 ($+0.6\%$)                                                \\
9   & 64     & 67.5       & 6500             & 6165 ($-5.2\%$)                                            & 6520 ($+0.3\%$)                                                \\ \bottomrule
\end{tabular}
\label{t:orders}
\end{table}

\begin{figure}[h]
\centering
\includegraphics[scale=0.6]{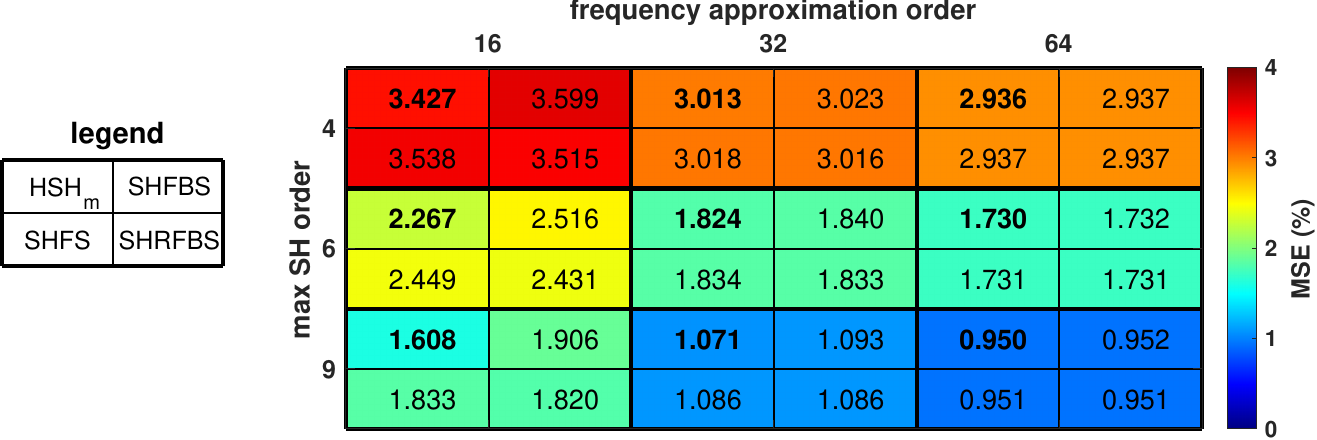}
\caption{Average MSE for investigated bases. Lowest value for each configuration is bolded.}
\label{f:tab}
\end{figure}

For all the configurations, MSE was determined for each of the 10 chosen subjects. Then, the MSE values were averaged and gathered in Fig.~\ref{f:tab}.

In each of the nine tested configurations of $L$ and $\eta$, the lowest average MSE was obtained for HSH$_{\text{m}}$. Out of the bases defined in SCS, SHRFBS yielded the best accuracy, followed by SHFS and then SHFBS. The differences between the bases in most configurations are very small, especially when higher $\eta$ are considered. The impact of basis choice rises for higher max SH orders, since the accuracy is to lesser extent limited by low-order SHs, which are a common element in all investigated bases.

Figure 2, besides comparing the accuracy for different bases, shows the importance of appropriate choice of approximation orders in space and frequency. For example, the 4:64 and 6:32 configurations both utilize approximately twice as much functions as the 4:32 configuration, but the gain in accuracy is not the same. In the case of the 4:64 configuration, the error does not decrease substantially, since it is mostly caused by low max order of SHs, not by low frequency approximation order; further increasing $\eta$ is thus much less effective than increasing $L$ in this case. The optimal values of approximation orders are discussed in more detail in section \ref{s:optimal}.

\subsection{Analysis in frequency}
\label{s:frequency}

\begin{figure}
\centering
  \includegraphics[scale=0.6]{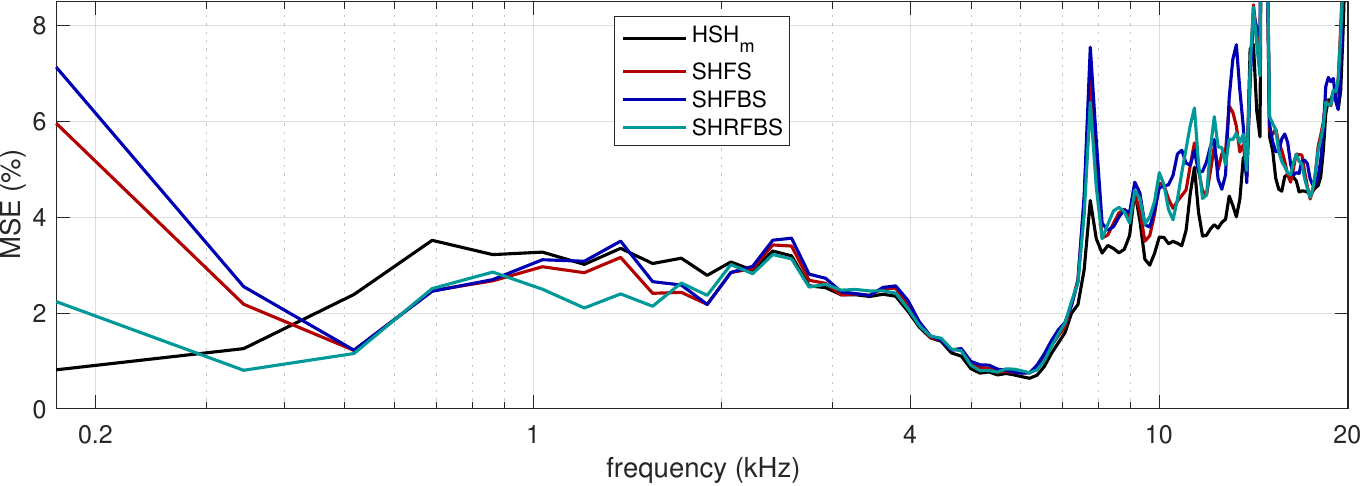}
  \caption{MSE in frequency for the 6:16 configuration for all analyzed representations}
  \label{f:freq}
\end{figure}

As the main difference between the discussed bases is the way they handle the frequency axis, they were first compared in the frequency domain. MSE was plotted for each of the representations for configuration 6:16 (Fig. \ref{f:freq}).

In the middle parts of the spectra, mean MSE is almost identical for all representations and the discrepancies can be observed mostly below or above the 2-7 kHz frequency region. For the higher frequencies, there are many variations, but overall accuracy level is similar across all the bases except HSH$_{\text{m}}$, which performs slightly better thanks to larger effective $\eta$. On the other hand, in the lower frequency region, the HSH representation yields the largest errors except for the lowest samples; in HCS, they all lie close to each other on the hypersphere, while in SCS they are more independent and so the least-squares solver can more easily sacrifice them for better fit in neighbouring regions. SHFS and SHFBS maintain similar accuracy across the entire frequency range, while SHRBFS follows a slightly different pattern, offering lower MSE in some regions. Since SHRBFS consistently achieves better results than the remaining functions defined in SCS, the latter are excluded from further analysis and the focus is put on the differences between the HSH$_{\text{m}}$ and SHRFBS representations as the two most accurate ones.

\subsection{Impact of approximation order change}
\label{s:change}

\begin{figure}
\centering
\subfigure[]{
  \includegraphics[scale=0.6]{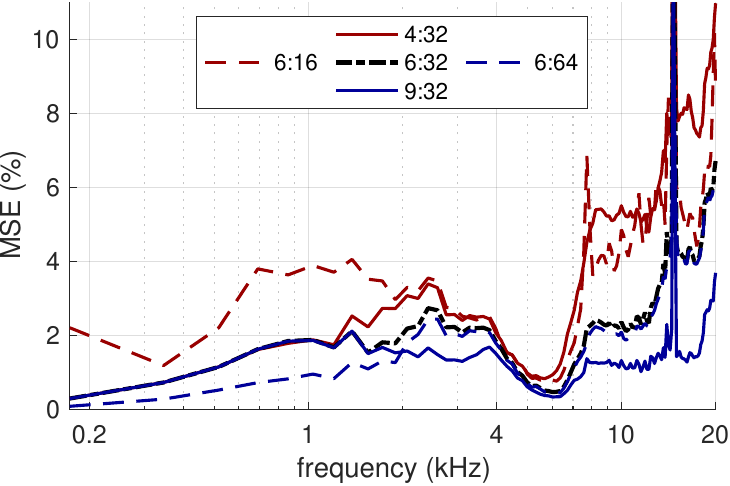}
  \label{f:frequencyHSH}}
\quad
\subfigure[]{
  \includegraphics[scale=0.6]{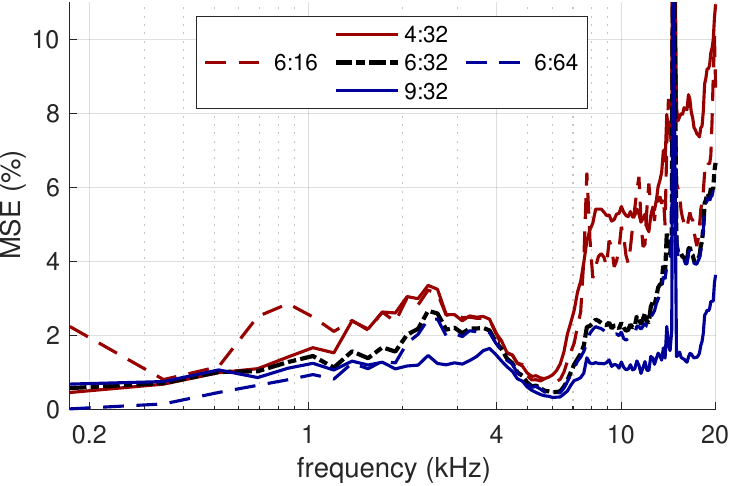}
  \label{f:frequencySHFS}}
\caption{MSE in frequency for different configurations for a) HSH$_{\text{m}}$ and b) SHRFBS. The same color indicates approximately the same number of functions.}
\label{f:change}
\end{figure}

In order to visualize the impact of changing $L$ and $\eta$, MSE was plotted in frequency for the 6:32 (the middle) configuration as well as configurations with only either $L$ or $\eta$ changed (Fig. \ref{f:change}).

In general, increased $L$ yields better accuracy in high frequencies while $\eta$ improves approximation more effectively in the lower frequency range. The latter is consistent with the theory, which suggests that directivity patterns in high frequency require higher max SH orders in order to be accurately expressed \citep{Zhang2010}. By proper balancing of $L$ and $\eta$, a more or less uniform accuracy across the entire frequency range can be achieved; although elevation cues lie mostly in the high frequency region, the lower frequency part of HRTFs are also important, as they are responsible for front-back direction discrimination \citep{Asano1990}.

\subsection{Analysis in direction}
\label{s:direciton}

\begin{figure}
\centering
\subfigure[]{
  \includegraphics[scale=0.6]{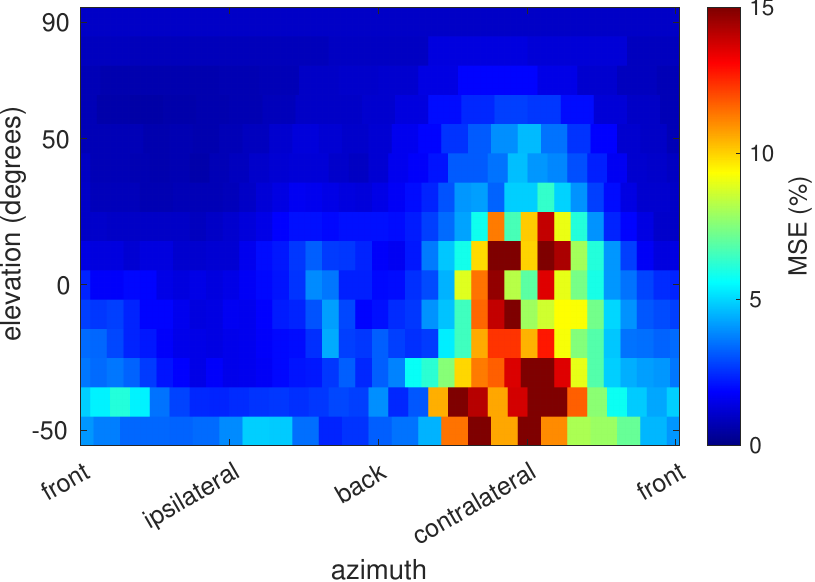}
  \label{f:dirHSH}}
\quad
\subfigure[]{
  \includegraphics[scale=0.6]{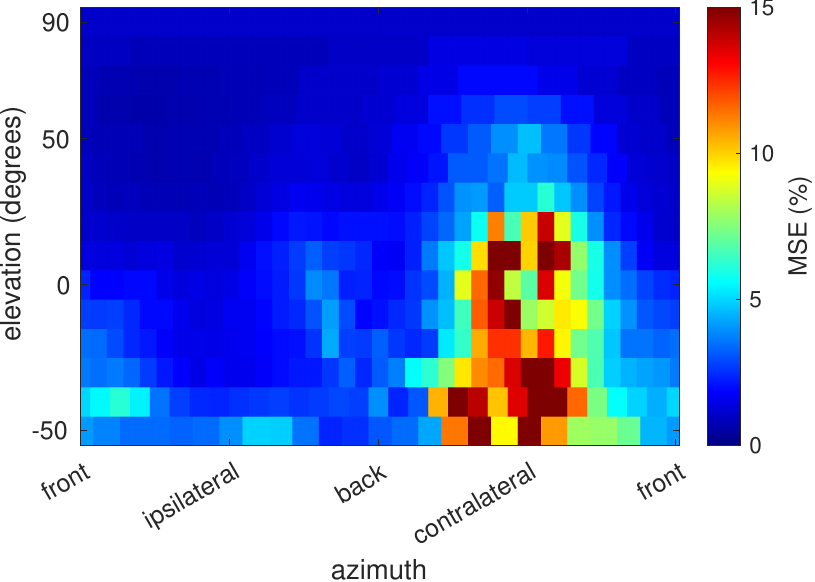}
  \label{f:dirSHFS}}
\caption{MSE in direction for the 6.32 configuration for (a) HSH$_{\text{m}}$ (b) SHRFBS}
\label{f:dir}
\end{figure}

Even though in each of the analyzed 4D representations the direction dependence is described by SHs, due to the space-frequency coupling, the choice of frequency representation directly impacts the accuracy in both frequency and space. Since for HRTFs the directions in azimuth matter with regard to the position of ear, the data for right ears were flipped and analyzed together with the data for left ears. Figure~\ref{f:dir} presents MSE depending on the relative direction of incoming sound for the 6:32 configuration for HSH$_{\text{m}}$ and SHRFBS.

\begin{figure}
\centering
  \includegraphics[scale=0.6]{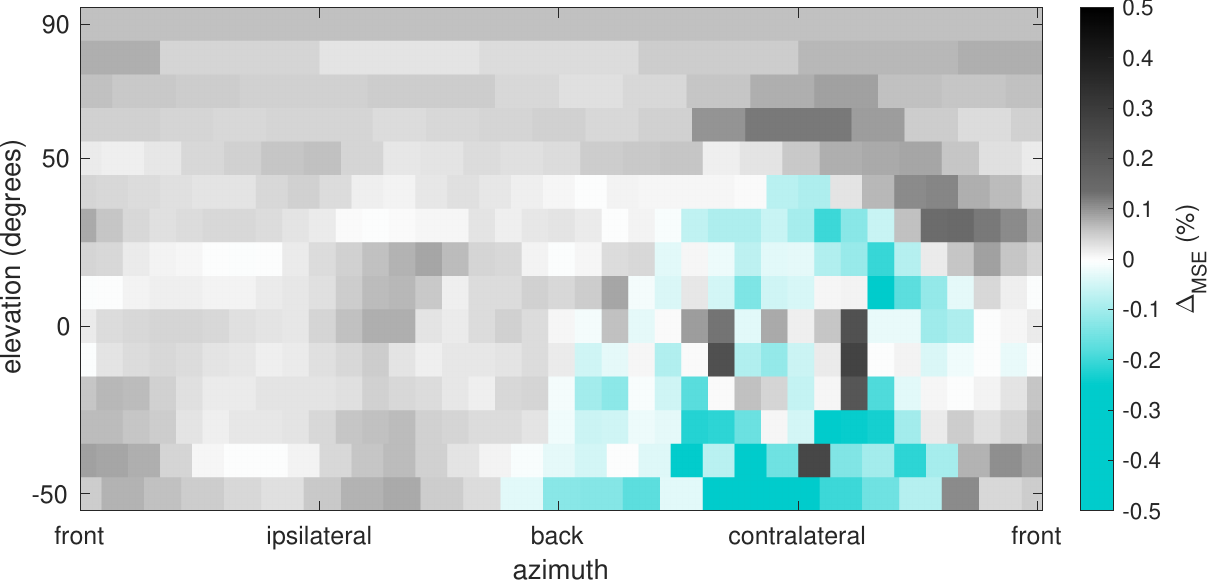}
  \caption{Difference between data from \ref{f:dirHSH} and \ref{f:dirSHFS}. Negative values show regions where the error is lower for SHRFBS, while positive values suggest better accuracy of HSH$_{\text{m}}$.}
  \label{f:dirDiff}
\end{figure}

In both cases, the largest errors occur on the contralateral side (i.e. on the left side for right ears and on the right side for left ears), especially at low elevations. Modeling of contralateral parts of HRTFs has been proven to be more difficult in both space and in frequency, as well as in compound representations (e.g. \citep{Kulkarni2004,Chen1995,Zhang2015,Romigh2015}), so such results were expected. However, even though both maps in Fig.~\ref{f:dir} look similar, a clear trend can be noticed after plotting the difference (Fig.~\ref{f:dirDiff}); the HSH representation yields larger errors on the contralateral sides than the SHRFBS one, but provides slightly better accuracy everywhere else, so that the mean error is balanced out to almost equal. Such distribution is advantageous for the HSH representation, since the information from contralateral ear is far less important in the localization process \citep{Morimoto2001,Macpherson2007}.

\section{Optimal approximation order}
\label{s:optimal}

\subsection{Choice of HRTF database representative}
\label{s:representative}

\begin{figure}
\centering
\subfigure[]{
  \includegraphics[scale=0.6]{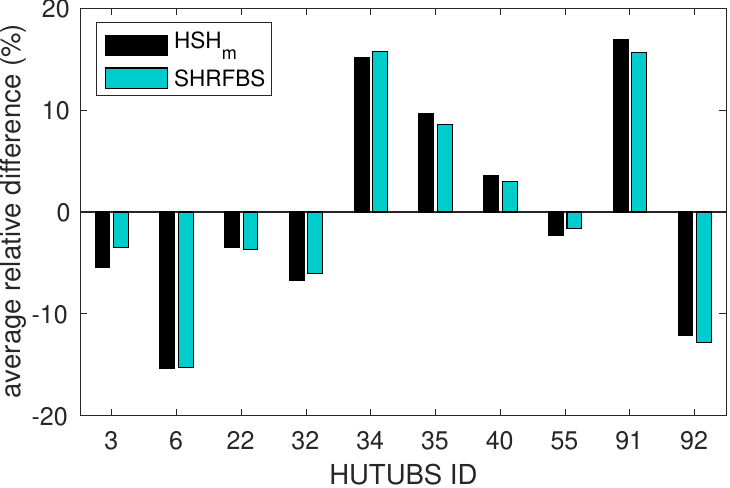}
  \label{f:repDiff}}
\quad
\subfigure[]{
  \includegraphics[scale=0.6]{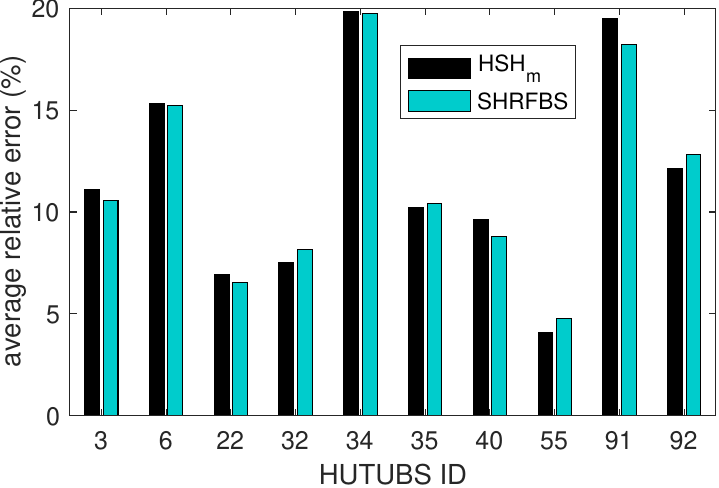}
  \label{f:repErr}}
\caption{Metrics used for the choice of the database representative for HSH$_{\text{m}}$ and SHRFBS: (a) ARD (b) ARE}
\label{f:representative}
\end{figure}

Since the deviation in results for different HRTFs from the same database is low, the following experiments were performed only on data for a single subject in order to reduce the required amount of computational resources. To chose the subject, two metrics were introduced: average relative difference (ARD) and average relative error (ARE) defined as:

\begin{equation}
\begin{aligned}
	\text{ARD}(s) &= \frac{1}{9}  \sum\limits_{L,\eta} \left( \frac{\text{MSE}(L,\eta,s) - \text{MSE}_{\mu}(L,\eta)}{\text{MSE}_{\mu}(L,\eta)} \right) \times 100\% \\
	\text{ARE}(s) &= \frac{1}{9}  \sum\limits_{L,\eta} \left( \frac{|\text{MSE}(L,\eta,s) - \text{MSE}_{\mu}(L,\eta)|}{\text{MSE}_{\mu}(L,\eta)} \right) \times 100\% \, ,
\label{e:AR}
\end{aligned}
\end{equation}

\noindent where the summing is performed over all configurations from Tab.~\ref{t:orders}, $\text{MSE}(L,\eta,s)$ is MSE for a given configuration and subject $s$, and $\text{MSE}_{\mu}(L,\eta)$ is $\text{MSE}(L,\eta,s)$ averaged over all subjects (as in Fig.~\ref{f:tab}). ARD indicates whether data for a given subject yielded on average lower or higher MSE than mean, while ARE determines average magnitude of the deviations from mean. ARD and ARE for all the subjects are presented in Fig.~\ref{f:representative}.

The differences in ARD and ARE for HSH$_{\text{m}}$ and SHRFBS are relatively low, even though the metrics' values vary substantially from subject to subject. This confirms that the evaluation can be performed quite reliably only on data for a single subject. Both ARD and ARE indicate that the accuracy of the HRTF with ID 55 is the closest to mean across all the configurations, and thus this HRTF was chosen to be a representative of the entire database in the following section. 

\subsection{Pareto frontier}
\label{s:pareto}

\begin{figure}[h]
\centering
\includegraphics[scale=0.6]{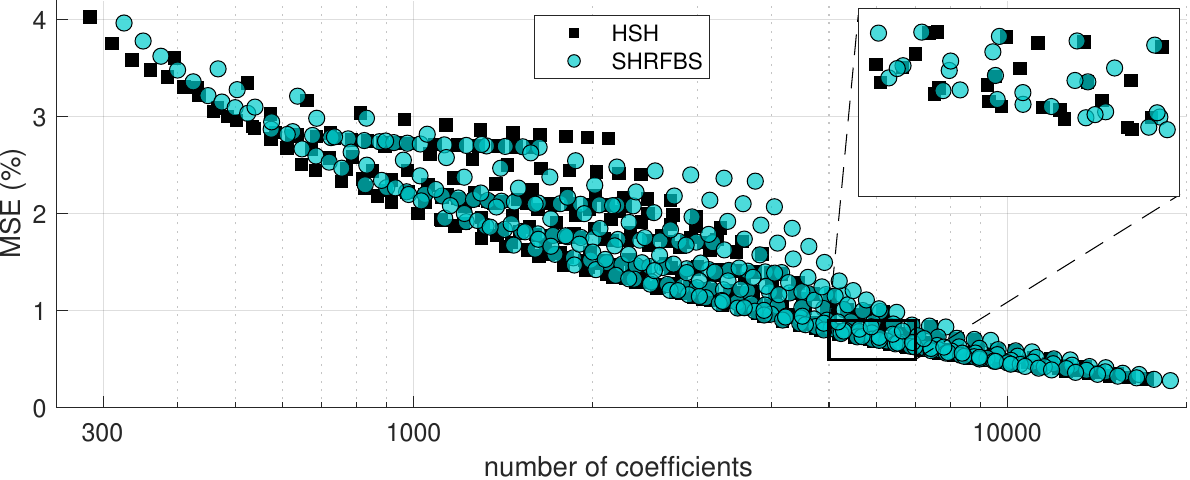}
\caption{MSE plotted against number of coefficients for a broad range of approximation orders (see text for details)}
\label{f:pareto}
\end{figure}

It is clear that the choice of $L$ and $\eta$ is crucial to the efficient application of the discussed HRTF representations. The question is particularly interesting as the investigated bases are defined in two different coordinate systems. The two basic criteria for an efficient HRTF representation are accuracy of data reproduction and number of coefficients (functions) required to achieve that accuracy. To solve such problems, the criteria are plotted against each other, resulting in so called Pareto frontier, i.e. set of optimal solutions. An optimal solution (also called undominated) is such that there exists no other solution which would improve on every criterion. MSE and $N_{\text{HSH}}$ or $N_{\text{SCS}}$ were determined for a broad range of approximation orders; $L$ from 4 to 16 (maximum possible $L$ declared by the HRTF database authors \citep{Brinkmann2019}) and $\eta$ from 12 to 64 with varying resolution. The results are gathered in Fig.~\ref{f:pareto}.

\begin{figure}
\centering
\subfigure[]{
  \includegraphics[scale=0.6]{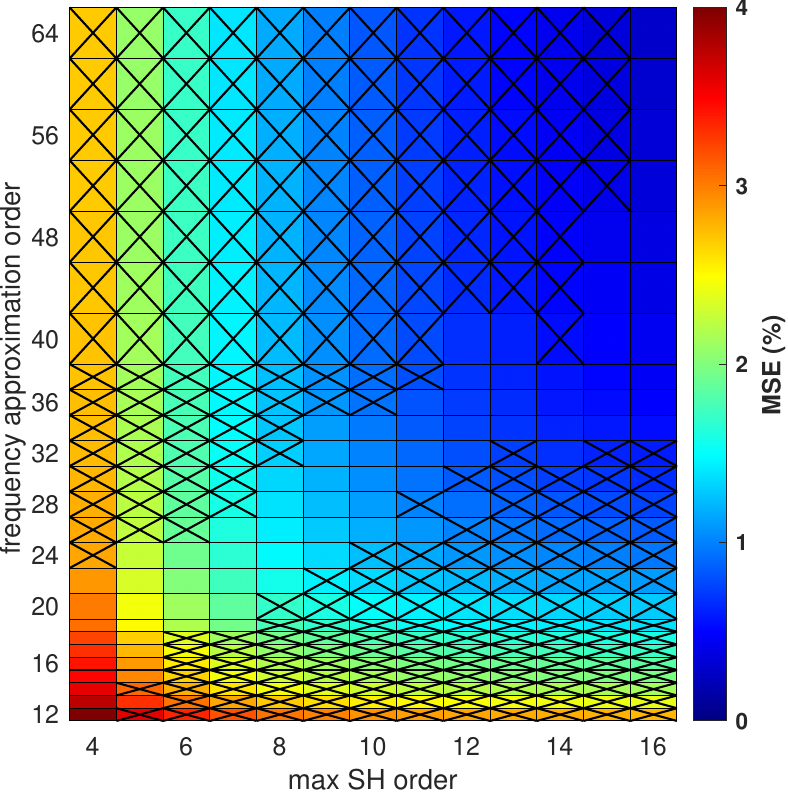}
  \label{f:optHSH}}
\quad
\subfigure[]{
  \includegraphics[scale=0.6]{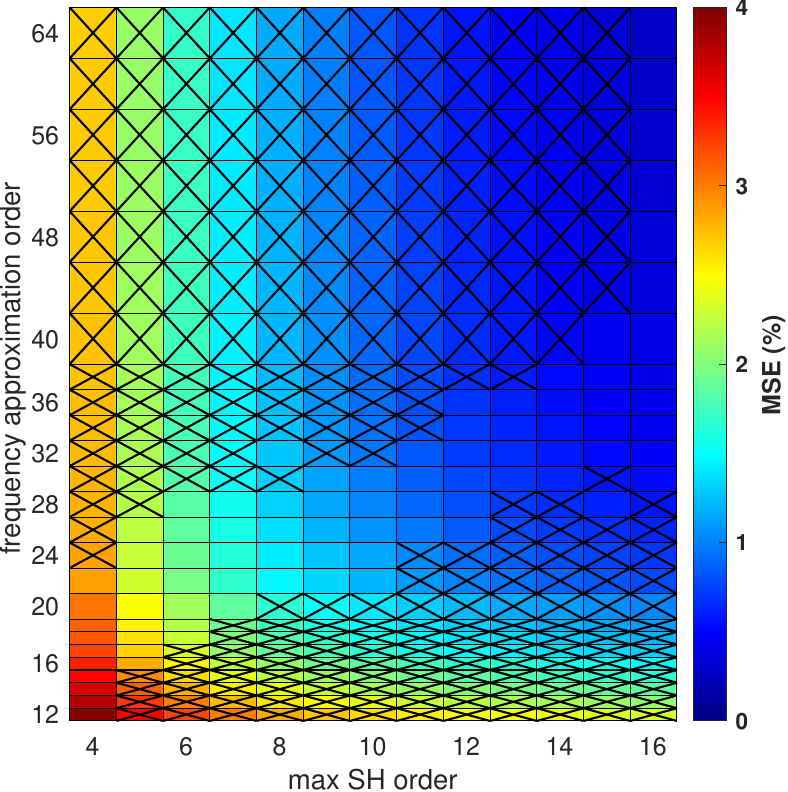}
  \label{f:optSH}}
\caption{MSE plotted by approximation orders: (a) HSH (b) SHRFBS. Crossed fields represent dominated configurations.}
\label{f:opt}
\end{figure}

The HSH and SHRFBS representations feature similar MSE to number of coefficient ratio, especially above 2000 coefficients\footnote{For reference, the closest undominated configurations to that threshold are 8.26 for HSH ($N_{\text{HSH}} = 1947$, $\text{MSE} = 1.423\%$) and 8.24 for SHRFBS ($N_{\text{SCS}} = 2025$, $\text{MSE} = 1.425\%$).}. Below that threshold, the HSH one dominates the SHRFBS approximations, although the differences are not large. However, it can also be noticed that some configurations are particularly inefficient, requiring much higher number of coefficients than needed to achieve the same accuracy. To further investigate this issue and find optimal values of $L$ and $\eta$, the MSE values for all configurations are grouped in Fig.~\ref{f:opt}.

Dominated configurations occur mostly when either $L$ or $\eta$ is disproportionately large. In order to maximize the efficiency of used basis, $\eta$ should be approximately three times as high as $L$. Some of the configurations marked as optimal in Fig.~\ref{f:opt} would likely get dominated if a finer $\eta$ resolution was used; on the other hand, many configurations marked as dominated would still be fine in practical applications, since the relative differences in error and number of coefficients can be minuscule. Furthermore, for a different HRTF set or a different preprocessing, the exact values might change, so the above results serve only as an indicator what should be approximate relation between $L$ and $\eta$ to efficiently utilize the bases.

\section{Discussion}
\label{s:discussion}

Since HRTFs in principal vary along both space and frequency, treating both these dependencies uniformly seems natural. The approximation only in space or only in frequency can be seen as suboptimal, as it means disproportionately large $\eta$ or $L$, respectively; increasing either of the two, the model accuracy converges to the accuracy of approximation only in the other dimension, i.e. MSE for low $L$ and high $\eta$ is almost the same as for an SH approximation (discrete in frequency) of the same $L$. However, even using 4D models with such disproportionate orders might be advantageous, as it provides continuity in frequency, allowing e.g. for reading HRTFs of any requested resolution (or, equivalently, reading HRIRs of any length).

The 4D approach features significant flexibility in terms of balancing accuracy and number of coefficients. Knowing the impact of changing approximation orders (see section \ref{s:change}), one can control not only the mean error, but also its levels in different regions, e.g. putting more focus on higher frequencies by increasing $L$ while keeping relatively low $\eta$. Even though optimal range of $L$ to $\eta$ ratio was determined for both HSHs and SHRFBS, it was based on MSE, which is a rather simplistic measure of accuracy and does not perfectly reflect perceived quality of approximation. However, listening tests are not feasible for such large number of tested configurations, and, what is more, the reliability of this method is relatively low; utilized in this research MSE appears to be a better predictor of subjective preference than a subjective rating issued by a different individual \citep{SzwajcowskiICA}. 

One of the challenges in continuous 4D models is allowing for higher spatial variability at high frequencies, which in discrete SH applications can be implemented as increasing max SH order. It is notable that out of HSHs, SHFS, SHFBS and SHRFBS, the best accuracy was obtained for the bases which accounted for that effect, either by mapping the HRTF data on the hypersphere or by representing the frequency axis with reversed Fourier-Bessel series, which natively provides higher magnitudes for higher values of arguments. Out of these two, the HSH representation has proved to be more efficient, particularly for low number of functions. Furthermore, HSHs seem to be better fit to preserve the ipsilateral parts, which are crucial for elevation perception \citep{Morimoto2001,Macpherson2007}.

The obvious application of the 4D models is data compression, in which case the HSH representation is superior to the ones defined in SCS, although not by a big margin. Raw HRTF dataset consists of 440 HRIRs, each of 256 samples, yielding 112 640 values in total. Assuming required reproduction error below 1\%, only 3795 HSH coefficients would suffice, meaning the compression ratio of around 30; for 2\% MSE, that value rises above 100. However, these numbers concern only the magnitude part and could be more or less lower depending on the complexity of phase modeling. Such compact description of HRTFs can be also beneficial as parametrization method, e.g. for machine learning purposes. Further reduction of data size could be achieved by using only a subset of coefficients \citep{SzwajcowskiISAV}.

While the HSH representation is slightly more efficient in terms of accuracy obtained for a given number of coefficients, the bases defined in SCS have simpler definitions, which could be useful in real-time computing, i.e. in binaural processing. In such applications, it might be worth to trade slightly lower reproduction accuracy for an easier generation and handling of the basis functions.

Computing 4D HRTF representations using the approach discussed within this paper relies on creating a large matrix $\mathbf{X}$, which consists of 4D basis function values for each frequency-direction sample and then solving Eq. \eqref{e:matrixLS} in the least-squares sense. Depending on the dataset resolution and on desired approximation orders, handling such matrix could require significant amount of memory and processing power. For the configurations discussed in this paper, the number of elements in $\mathbf{X}$ ranges from 32 102 400 (HSH, $L = 4, \eta = 12$) to 2 115 942 400 (SHRFBS, $L = 16, \eta = 64$), which, using the default \textsc{Matlab} double-precision format (64 bits per element), results in up to about 13.5 GB of RAM needed to handle the matrix. Application of this method on a dataset with fine resolution and/or for large approximation orders could thus be out of reach of a typical PC. However, very high precision is excessive from the perceptual point of view and, for smaller approximation orders, a high-resolution HRTF dataset could be downsampled before least-squares computation without decreasing the accuracy significantly. Once the coefficients $\alpha_{nl}^{m}$ are determined, decoding back to the frequency-space domain is a simple task and requires much less computational resources.

All the models in this paper were discussed in the context of HRTFs; however, their structure allows for representing any kind of directivity data, such as e.g. sound source directivity. Since they pose an alternative to SH approximations, their scope of applications and possibilities regarding different approaches to 4D models are very wide.

\section{Conclusions}
\label{s:conclusions}

Within this work, the theory and experimental results were provided for HRTF models based on 4D basis functions: HSHs, SHFS, SHFBS and SHRFBS. Out of these four, HSHs appeared to be the best fit to represent HRTF data, while the other three displayed similar levels of efficiency for this task, but with clear order from best to worst: SHRFBS, SHFS and SHFBS. The models were compared both in the frequency and in the space domain, putting extra focus on the impact of chosen approximation orders. Detailed study of mean errors for a given number of functions revealed that the HSH representation is superior to the SHRFBS one when less than 2000 functions are used for approximation and approximately equally as accurate above that threshold. Furthermore, range of optimal orders was investigated, suggesting that frequency approximation order should be 2 to 4 times higher than max SH order.

The discussed models can be applied to a wide range of directivity functions, including HRTFs, sound source directivity, microphone directivity, etc., providing robust parametrization and compression method. The representations are more efficient than previously studied approximations only in either space or frequency, and, by treating these two dependences uniformly, inherent coupling between them is maintained. The downside of this method is that it requires significant amount of computational resources. However, with ever-growing computation industry, it can hardly be considered a drawback.

\section*{Acknowledgements}
This research was supported by the National Science Centre, project No. 2020/37/N/ST2/00122. The author would like to thank Anna Snakowska and Łukasz Gorazd for advice and helpful discussions as well as anonymous reviewers who provided important insight at different stages of the project.



  \bibliographystyle{elsarticle-num} 
  \biboptions{sort&compress}
  \bibliography{biblio}

%
%
%
%
\end{document}